\documentclass[conference]{IEEEtran}
\IEEEoverridecommandlockouts
\usepackage{cite}
\usepackage{amsmath,amssymb,amsfonts}
\usepackage{algorithmic}
\usepackage{graphicx}
\usepackage{textcomp}
\usepackage{xcolor}
\usepackage{algorithm}
\usepackage{color,soul}
\usepackage{subcaption}
\usepackage{booktabs, makecell,tabularx}

\def\BibTeX{{\rm B\kern-.05em{\sc i\kern-.025em b}\kern-.08em
    T\kern-.1667em\lower.7ex\hbox{E}\kern-.125emX}}
\begin{document}

\title{Creating an Explainable Intrusion Detection System Using Self Organizing Maps}


\author{\IEEEauthorblockN{Jesse Ables\IEEEauthorrefmark{1}, Thomas Kirby\IEEEauthorrefmark{1}, William Anderson\IEEEauthorrefmark{1}, \\ Sudip Mittal\IEEEauthorrefmark{1}, Shahram Rahimi\IEEEauthorrefmark{1}, Ioana Banicescu\IEEEauthorrefmark{1}, and Maria Seale\IEEEauthorrefmark{2}}

\IEEEauthorrefmark{1} Department of Computer Science \& Engineering \\ Mississippi State University, Mississippi, USA, \\(email: \{jha92, tmk169, wha41\}@msstate.edu, \{mittal, rahimi, ioana\}@cse.msstate.edu)\\
\IEEEauthorrefmark{2} U.S Army Engineer Research and Development Center \\ Vicksburg, Mississippi, USA, (email: maria.a.seale@erdc.dren.mil)

}

\maketitle

\begin{abstract}


Modern Artificial Intelligence (AI) enabled Intrusion Detection Systems (IDS) are complex black boxes. This means that a security analyst will have little to no explanation or clarification on why an IDS model made a particular prediction. A potential solution to this problem is to research and develop Explainable Intrusion Detection Systems (X-IDS) based on current capabilities in Explainable Artificial Intelligence (XAI). In this paper, we create a Self Organizing Maps (SOMs) based X-IDS system that is capable of producing explanatory visualizations. We leverage SOM's explainability to create both global and local explanations. An analyst can use global explanations to get a general idea of how a particular IDS model computes predictions. Local explanations are generated for individual datapoints to explain why a certain prediction value was computed. 
Furthermore, our SOM based X-IDS was evaluated on both explanation generation and traditional accuracy tests using the NSL-KDD and the CIC-IDS-2017 datasets. 

\end{abstract}


\section{Introduction}
The use of Artificial Intelligence (AI) in cyber-defense solutions, particularly Intrusion Detection Systems (IDS), has been gaining traction to protect against a wide range of cyber attacks. While these AI models have high detection rates, high false positive and false negative rates can dissuade a security analyst from using an AI enabled IDS 
\cite{Marshan}. 
These IDS built using AI and deep learning methods are black boxes, meaning a security analyst will have little to no explanations and clarifications on why a model made a particular prediction. With the rise in cyber attacks on critical infrastructure, government organizations, and business networks, there is a pressing need for an explainable, automated detection system that can provide real-time aid to an analyst. 

{Intrusion Detection Systems are generally utilized as part of a larger cybersecurity defense effort at an organization generally located in a Cyber-Security Operations Center (CSoC).} These systems monitor networks and automate attack detection by comparing network activity to the signature of known attacks or by detecting behavior that is anomalous to benign network patterns \cite{raytheon_2017}. Through these methods, a security analyst can use an IDS to detect improper use, unauthorized access, or the abuse of a network. Analysts can then create mitigating strategies to minimize damages and costs of the malicious behavior. The usefulness and cost effectiveness of IDS have therefore been the subject of much research \cite{belouch2018performance,wu2010use}. 

Previous work in AI enabled IDS has generally focused on improving detection rates while limiting false positives and false negatives. These techniques have been effective at achieving high detection rate, but have failed to provide explanations for their computed predictions. Without the ability to understand how a model reached a decision and which features were relevant to the decision computation, a security analyst will give less credence to these AI enabled IDS. Opaque Deep Learning methods in particular, can be considered as black boxes providing no explanations and feature relevance information, severely limiting adoption in real world cyber-defense scenarios \cite{lipton2018mythos}. 

A potential solution to this problem is to research and develop Explainable Intrusion Detection Systems (X-IDS) based on current capabilities in Explainable Artificial Intelligence (XAI) \cite{subashxids}.
The guidelines proposed by the Defense Advanced Research Projects Agency (DARPA) indicate that to be explainable, an AI should explain the reasoning for its decisions, characterize its strengths and weaknesses, and convey a sense of its future behavior \cite{gunning2019darpa}. An X-IDS that is transparent in its behavior and decision making process, will empower a security analyst to make better informed actions, understand the feature composition of a prediction, help CSoCs defend from known attacks, and quickly understand zero-day attacks. To address this need, we create an X-IDS using Self Organizing Maps (SOMs).

Data collected from modern networks contain potentially hundreds of different features about the traffic flow, operating systems, network protocols, and other metadata. SOMs work by representing this high dimensional data on a 2-dimensional plane. This also includes maintaining the topographical relationship of the data by grouping similar data \cite{kohonen1982self}. Through this dimensional reduction and various other SOM visualization techniques, a security analyst can view both global and local explanations about a potential attack rather than an opaque prediction generated by a black box model. 

As the need for explainable cyber-defense systems increases and to address the lack of XAI research in the field of IDS, the main objective of this paper is to demonstrate the explainability of the SOM based X-IDS rather than creating the most accurate system. Higher accuracy systems can be developed by using complex derivative architectures. However, further research is necessary to make them explainable. Our goal in this paper is to increase trust in IDS and help CSoCs defend from attack through the use of explainable insights. As a secondary focus, we also provide the accuracy scores of our SOM based X-IDS system trained on the NSL-KDD and CIC-IDS-2017 datasets. 

Major contributions presented in this paper are -
\begin{itemize}
    \item A SOM based X-IDS, built using DARPA's proposed guidelines for an explainable system. This system is able to produce robust, explanatory visualizations of the SOM model and create accurate IDS predictions.
    \item A Local and Global explainability analysis using the SOM explainable architecture. The explanation module creates a collection of explainable visualizations that can be used by a security analyst to understand predictions.
    \item A performative analysis using NSL-KDD and CIC-IDS-2017. The SOM based model is able to achieve accuracies as high as 91\% on NSL-KDD and 80\% on CIC-IDS-2017 datasets.
\end{itemize}

\begin{figure*}[htbp]
    \centering
    \includegraphics[scale=.7]{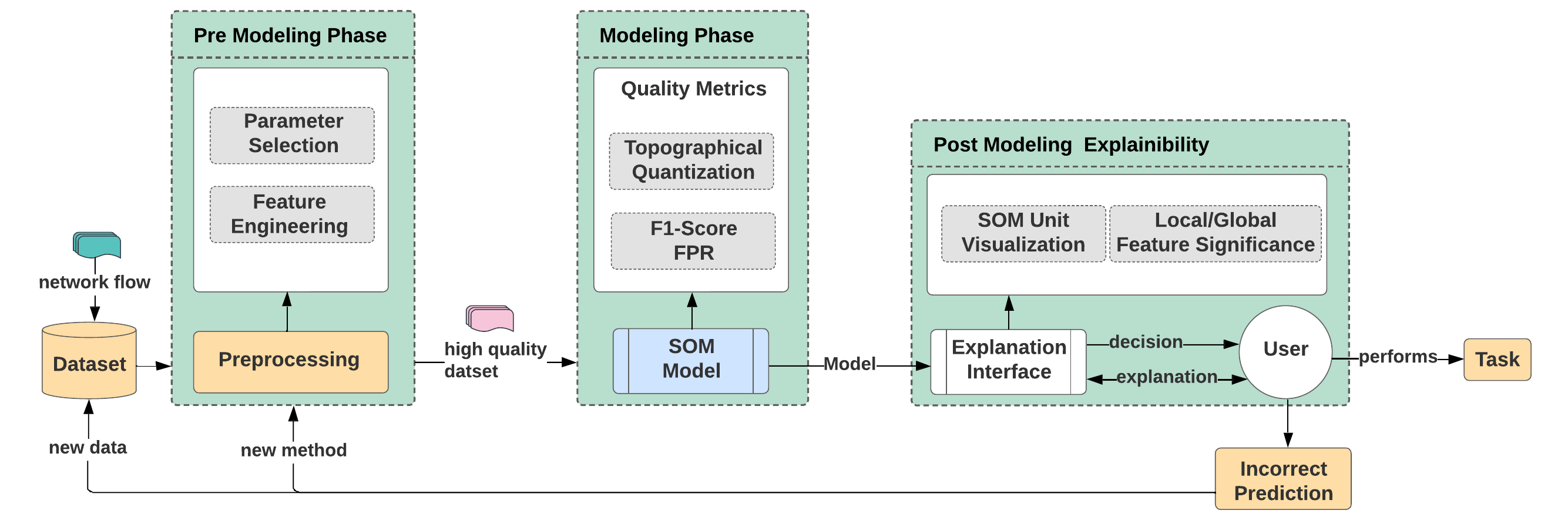}
    \vspace*{-3mm}
    \caption{Architecture for an Explainable Intrusion Detection System (X-IDS) utilizing Self Organizing Maps (SOMs), based on DARPA's recommended architecture for Explainable Artificial Intelligence (XAI) systems \cite{gunning2019darpa}.}
    \label{fig: architecture}
\end{figure*}

The rest of the paper is outlined as follows - In Section \ref{relwork}, we discuss some related work on IDS, XAI, and X-IDS. Section \ref{X-SOM} briefly describes the SOM algorithm and how it can be used to achieve explainability. Section \ref{Architecture}, outlines our SOM based X-IDS with its architecture presented in Figure \ref{fig: architecture}. Section \ref{Results} lists our experimental results. Finally, the conclusion and future work has been presented in Section \ref{Conclusion}.

\section{Related Work}\label{relwork}

In this section, we present some related work on Intrusion Detection Systems (IDS), Explainable Artificial Intelligence (XAI), and Explainable Intrusion Detection Systems (X-IDS).

\subsection{Intrusion Detection Systems (IDS)}
An intrusion refers to an action that obtains unauthorized access to a network or system \cite{denning1987intrusion}. Intrusions can be characterized by a violation of Confidentiality, Integrity, or Availability (CIA). An IDS consists of tools, methods, and resources that help a CSoC protect an organization \cite{bace2001intrusion,mcdole2021deep}.

IDS can be classified as either a host-based IDS or network-based IDS. Host-based IDS are placed on a host system and monitor host activity, incoming and outgoing network traffic \cite{LetouHIDS}. Network-based IDS are built to survey and protect a network of hosts from intrusion \cite{Mukherjee1994NetworkID}. In addition, IDS can also be categorized into operation-based classes, such as signature, anomaly, and hybrid. Signature-based IDS operate by preventing known attacks from accessing a network. The IDS compares incoming network traffic to a database of known attack signatures. Notably, this method has difficulty in preventing \textit{zero-day} attacks \cite{sharma2014evolution}. Anomaly-based IDS look for patterns in incoming traffic to recognize potential threats and leverage complex AI models \cite{Chandola2009AnomalyDA,mcdole2020analyzing}. A significant drawback of this approach is the the tendency for such systems to categorize legitimate, unseen behavior as anomalous. Hybrid-based IDS incorporate the design philosophy of both signature-based and intrusion-based IDS to improve the detection rate while minimizing false positives \cite{szczepanski2020achieving, pang2021explainable}.

Current AI enabled anomaly-based IDS can be further divided into black box and white box models. White box models are considered \textit{easy to understand} by an expert. This allows the expert to analyze the decision process and understand how the model renders its decision. This "semi-transparent" property allows white box models to be deployed in decision sensitive domains, where auditing the decision process is a requirement. White box models may use regression-based approaches \cite{subba2015intrusion}, decision trees \cite{mahbooba2021explainable}, and SOMs \cite{langin2011annabell}. Black box models, on the other hand, have an opaque decision process. This opaqueness property makes establishing the relationship between inputs and the decision difficult, if not outright impossible. Black box models comprise nearly all the AI enabled state-of-the-art approaches for IDS, as the focus is traditionally on model performance, not explainability. Examples of black box models are Isolation Forest \cite{Liu2008IsolationF}, One-Class SVM \cite{Schlkopf1999SupportVM}, and Neural Networks \cite{ZhangNN}.


\subsection{Explainable Artificial Intelligence (XAI)}

As previously stated, state-of-the-art approaches for IDS, as well as machine learning as a whole, focus on model performance through the lens of model accuracy. This focus on model accuracy has driven the development further away from modeling approaches that are transparent or have methods of explainability. In turn, this creates a separation between model inference and \textit{understanding} model inference, which gives the inability to confirm model fairness, privacy, reliability, causality, and ultimately trust.


The notion of XAI dates back to the 1970s. Moore et al. \cite{moore1988explanation} surveyed works from the 1970s to the 1980s, detailing early methods of explanations. Some early explanations consisted of {canned text} and code translations, such as the 1974 explainer MYCIN \cite{shortliffe1974mycin}. We can find a more current definition of XAI by DARPA \cite{gunning2019darpa}. They define XAI as systems that are able to explain their reasoning to a human user, characterize their strengths and weaknesses, and convey a sense of their future behavior. In turn, the system offers some form of justification for its action, leading to more trust and understanding of the system. The explanations from an XAI system help the user not only in using and maintaining the AI model, but also helping users complete tasks in parallel with the AI system. Tasks can include doctors making medical decisions \cite{shortliffe1974mycin, holzinger2017we, lindsay2020explainable}, credit score decisions  \cite{chun2021study}, detecting counterfeit banknotes \cite{han2019joint} or CSoC operators defending a network \cite{gunning2019darpa, darpa2016broad}. 


\subsection{Explainable Intrusion Detection Systems (X-IDS)}
Explainable Intrusion Detection Systems (X-IDS) are still an emerging sub-genre in the field. The need for explainability in IDS is becoming increasingly necessary both to warrant further application in decision sensitive domains, as well as to supplement and empower existing knowledge techniques (e.g. data mining, rule-based development) that black boxes obfuscate. The users need to be confident in the predictions or recommendations computed by an IDS. {Understandable explanations allow users to perform their tasks correctly. The stakeholders of an IDS (e.g. CSoC operators, developers, and investors)} are individuals who will be dependent on the performance of the system. CSoC operators will be performing defensive actions based on prediction and explanation results. Developers can use explanations to fortify the model in areas where it is weak. Investors may need explanations to help them in making budgeting decisions for their company. 

The current literature consists of many different black box and white box models being used alongside explanation techniques. Common explainer modules for black box models are Local Interpretable Model-agnostic Explanations (LIME) \cite{ribeiro2016should}, SHapley Additive exPlanations (SHAP) \cite{lundberg2017unified}, and Layer-wise Relevance Propagation (LRP) \cite{binder2016layer}. Modern techniques for explaining black box models consist of creating surrogate models that generate explanations either locally or globally. Other methods involve propagating predictions backwards in a Neural Network or decomposing a gradient. More novel approaches have also experimented with making datasets explainable \cite{islam2019domain} or making GUIs for explainable systems \cite{wu2020feature}.  

\section{Explainable Self Organizing Maps}
\label{X-SOM}

In this section, we briefly describe the theoretical and the practical aspects of SOMs and how they can be utilized to detect intrusions and generate explanations.

\subsection{Self Organizing Maps (SOMs)}

Self Organizing Maps (SOMs), sometimes referred to as Kohonen Maps \cite{kohonen1982self,oja1999kohonen}, Kohonen Self Organizing Maps \cite{guthikonda2005kohonen}, or Kohonen Networks \cite{kohonen2007kohonen}, are a class of unsupervised machine learning algorithms. SOMs are comprised of a network of individual units, each of which has a feature vector of the same size as the dimension of training data. Some implementations also include an \textit{(x,y)}  coordinate to allow unit movement in a two-dimensional (2D) space. This 2D space is typically represented as a square or a hexagonal grid, to more easily visualize the represented space. 

Training a SOM model, outlined in Algorithm \ref{alg:cap}, utilizes the following steps: First, a random training sample is picked. Then, the Best Matching Unit (BMU) is calculated by finding the smallest euclidean distance from the training sample to a SOM unit. After the BMU is found, it and its neighbors are updated using the formula $w_i = w_i - \lambda * (w_i - i_i)$, where $w$ is the set of BMU weights and $i$ is the set of feature values. $\lambda$ {is the learning rate function that considers the current training iteration and distance from the BMU.} Lastly, the learning rate, neighborhood radius, and current iteration numbers are updated. {The function} $\lambda$ {works in a way that it decreases during the course of the training process.}

 \begin{algorithm}
 
 \caption{SOM Algorithm}
  \label{alg:cap}
 \begin{algorithmic}[1]
 \renewcommand{\algorithmicrequire}{\textbf{Input: }}
 \renewcommand{\algorithmicensure}{\textbf{Output:}}
 \REQUIRE n, m, T
 \ENSURE  W
 \\
  \STATE Allocate n * m element array W
 \\
  \FOR{each node in W}
  \STATE Allocate N element array with random values [0,1]
  \ENDFOR
  \WHILE {$t < T$}
  \STATE Pick a training sample
  \STATE Find Best Matching Unit using Euclidean Distance
  \STATE Update BMU elements: $w_i=w_i-\lambda * (w_i - i_i)$
  \STATE Update BMU Neighbors
  \STATE Update Learning Rate and Radius
  \STATE $t += 1$
  \ENDWHILE
 \RETURN $W$ 
 \end{algorithmic} 

 \end{algorithm}



SOMs have some unique advantages that come with their application. The first is algorithmic simplicity. As shown in Algorithm \ref{alg:cap}, the brevity of the algorithm helps to maintain the desired properties of algorithmic decomposability and tractability. Additionally, due to its unsupervised nature, SOMs can work on a variety of datasets and applications (e.g. data mining and discovery), not just prediction \cite{Ong1999DataMU}. By design, SOMs convert high-dimensional data into a lower dimensional representation. This representation can be topologically clustered and explained through visualizations \cite{pachghare2009intrusion}. One challenge that comes with the application of SOMs is the selection of the size parameter, as the size does not dynamically adjust and there is no \textit{best size} heuristic \cite{Breard2017}. Finally, another challenge with SOMs is their scalability, both in their time complexity, $O(N^2)$, and space complexity. More methods, such as those in \cite{liu2018scalable}, are needed to address these challenges.

\subsection{SOMs and Intrusion Detection}

In the past, SOMs have been used to create many IDS. These studies focused on building accurate IDS and did not discuss explainability. Among these approaches, SOMs were used to create both host-based \cite{lichodzijewski2002host} and network-based \cite{de2015implementation, pachghare2009intrusion, rhodes2000multiple} IDS. The majority of these methods simply trained a SOM based IDS and illustrated mappings between datapoints and the associated BMU. The approaches in \cite{albayrak2005combining,rhodes2000multiple} use multiple SOMs in conjunction with one another to create a more effective IDS. Only one approach \cite{de2015implementation} discussed the false positive rate and accuracy of a SOM-based IDS. Their method for prediction involved assigning a label to BMUs based on the training dataset. Using this approach meant that not all SOM units were assigned a label. The authors utilized Gaussian Mixture Modeling (GMM) to make predictions when a testing sample was similar to an unlabeled unit.

\subsection{SOMs and Explainability}
Once trained, SOMs are able to illustrate mappings between datapoints and the associated BMU. As this is generally a 2D representation of the feature space, it can be visually understood by the user. This advantageous SOM property makes them \textit{explainable}. SOM's explainablity can be divided into \textit{Global} and \textit{Local} explanations. 

Global explanations are used to give a general idea of how a particular model computes predictions. The U-Matrix (See Section \ref{pme}), is a commonly used technique \cite{ultsch-kohonens-data-analysis-1990}. This additive metric works by summing the distance to each of a unit's neighbors. A group of low scores represents clusters in the map, while a group of high scores signifies sparseness. The Starburst U-Matrix is an updated variation of the U-Matrix. This updated version helps visualize cluster sizes and locations on the map. Other clustering representations, like K-means clustering, can also aid in visualization. For more fine-grained data, feature heat-maps can be created to visualize SOM feature values and their importance. These techniques provide global explanations.

Local interpretations of data are a more recent development for explaining SOMs. These explanations are generated for individual sample datapoints and are used to explain why a certain prediction value was computed. This allows the user to understand the decision process of the SOM model. The primary use of this method is to explain and visualize feature importance. When making a prediction, a datapoint's features are scored based on how impactful they are to the computed prediction. Wickramasinghe et al. \cite{Wickramasinghe2021} developed an explainable SOM for Cyber-Physical Systems. Their system created both local and global explanations by data-mining a SOM model. The mined information was used to create visualizations including histograms, T-distributed Stochastic Neighbor Embedding (t-SNE), heat maps, U-Matrix, component planes, and U-Map. This variety of visualizations allow the SOM to be explainable not only to domain experts, but also non-domain experts.



\section{X-IDS Architecture}
\label{Architecture}

An X-IDS's main goal is to help stakeholders protect their networks and understand various relevant events. The system should act as both a guard and adviser for network security. When an IDS discovers an intrusion, the user should be notified to prevent a possible attack. Explanations generated by the X-IDS should assist CSoC operators in their mission to protect their organization. 
To help achieve this goal, we propose the proof of concept SOM based X-IDS architecture in Figure \ref{Architecture}. The proposed architecture is based on DARPA's recommended architecture for XAI systems \cite{gunning2019darpa}. Components of the framework can be changed to suit users' needs. The architecture is abstract enough that methods other than SOMs can be interchanged to create different X-IDS. The architecture consists of three stages: pre modeling, modeling, and post modeling explainability. In the first phase, the model preprocesses raw network data captured into high quality datasets, and selects parameters for the SOM model. Next, the model is trained during the modeling phase. Metrics are recorded to determine the newly trained model's quality. Lastly, in the post modelling phase, the SOM is data-mined to generate explanatory visualizations that allow users to understand how predictions are generated. In the next subsections, we describe each of these phases in detail.

\subsection{Datasets and Pre Modeling Phase}\label{data}

In this work, NSL-KDD \cite{Tavallaee2009} and CIC-IDS-2017 \cite{Panigrahi2018} were used to test the explainability and effectiveness of our architecture. NSL-KDD was chosen because of its wide use in the literature. It allows our method to be compared to other existing IDS. CIC-IDS-2017 includes more modern attacks and is useful for testing an unbalanced dataset. The datasets are passed through a preprocessing module that normalizes the data. Additionally, the architecture uses Bayesian Probability of Significance \cite{Hamel2012a} to select features. Any feature significance value over a designer selected threshold is included in the preprocessed dataset. The resulting high quality dataset is used during the modelling phase.  

\subsection{Modeling Phase}

The modeling phase begins by training the SOM model using the high quality dataset generated during the pre modeling phase. For this paper, we use the POPSOM implementation \cite{yuan2018implementation}. Training parameters include total training iterations, learning rate, and SOM size. At the end of the training session, the model will be tested to produce topographical error, quantization error, F1-score, precision, recall, and a confusion matrix. The confusion matrix can be used to determine both the false positive and false negative rate for the model. The quality metrics are used to determine if a model has been sufficiently trained to generate explanations.

\subsubsection*{Quality Metrics}

There have been various metrics and measures proposed to evaluate the quality of a trained SOM. These include quantization error, topographic accuracy, embedding accuracy, and convergence index. Quantization error was used by Kohenen \cite{Kohonen1998}, and measures the average distance between nodes and the data points. To measure how much the features of the input space have been preserved in low dimensional output space, a topographic error is used. The topographic error is measured by evaluating how often the BMU and second BMU are next to each other \cite{Breard2017},\cite{Lampinen1992}. Map embedding accuracy is similar to quantization error and it measures how similar the distribution of the input data is with respect to that of the SOM units \cite{Hamel2016}. In order to measure both topographic preservation and distribution similarity between the input and SOM units, the convergence index was proposed to be a measure that linearly combines map embedding accuracy and topographic accuracy \cite{Tatoian2018}. Prediction accuracy metrics are also important to include in an IDS architecture. These metrics include F1-score, false positive rate, and false negative rate. These measurements allow the architecture to be compared to other existing IDS.

\label{local-global}
\begin{figure*}
\label{fig: local-global}
     \begin{subfigure}[b]{0.38\textwidth}
         \hspace{15pt}
         \subcaptionbox{NSL-KDD Local Anomaly Explanation\label{fig: nsl-local}}{%
         \includegraphics[scale=.6]{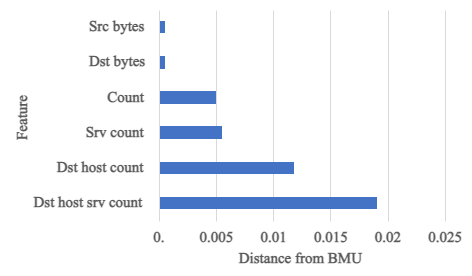}}
         
     \end{subfigure}
     \hspace{70pt}
     \begin{subfigure}[b]{0.3\textwidth}
         \subcaptionbox{NSL-KDD Global Feature Significance\label{fig: nsl-global}}{%
         \includegraphics[scale=.6]{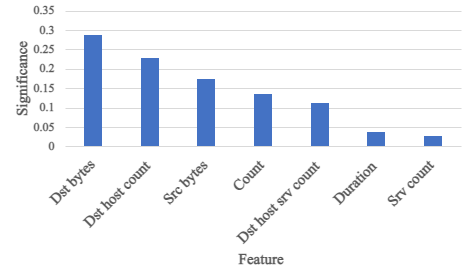}
         }
     \end{subfigure}
     \vfill
     \begin{subfigure}[b]{0.3\textwidth}
     \subcaptionbox{\mbox{CIC-IDS-2017 Local Normal Explanation}\label{fig: cic-local}}{%
         \includegraphics[scale=.7]{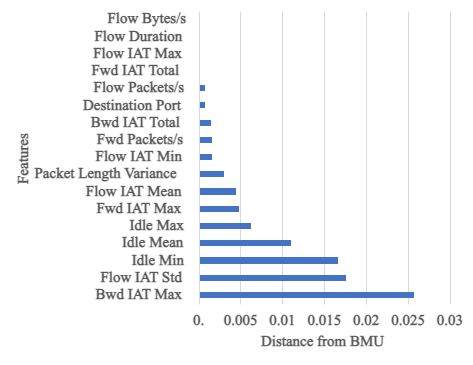}
         }
         
     \end{subfigure}
     \hspace{80pt}
     \begin{subfigure}[b]{0.3\textwidth}
         \subcaptionbox{\mbox{CIC-IDS-2017 Global Feature Significance}\label{fig: cic-global}}{%
         \includegraphics[scale=.8]{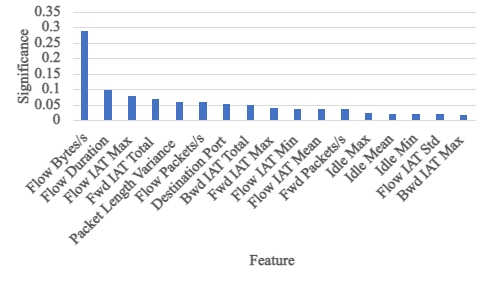}
         }
     \end{subfigure}
        \caption{These figures show the local and global feature explanations for both the NSL-KDD and CIC-IDS datasets. (a) The local explainability of a prediction is defined by the distance between feature value and BMU. More important features have a lower score than less important features. This figure shows the feature importance for an anomalous sample from the NSL-KDD testing set. (b) Global feature significance is calculated using Bayesian Probability of Significance \cite{Hamel2012a}. Higher values are considered more important than lower values.}
        
\end{figure*}

\subsection{Post Modeling Explainability}\label{pme}

Once the modeling phase has been completed and quality metrics have ensured that the model is a good representation of the data, the model can be used to perform a variety of explainability and visualization tasks. The model itself is a list of SOM units and the weights associated with these units. Visualization tasks include creating local and global explanations, a U-Matrix, and feature heatmaps.

\subsubsection{Local and Global Explanations}
    
Global and local interpretability can be achieved by examining important features of the trained SOM, and utilizing this information to generate an explanation for a specific data instance classification or cluster classification \cite{wickramasinghe2021explainable}. Global significance for NSL-KDD is shown in Figure \ref{fig: nsl-global} with higher values denoting that a feature has a higher probability of being important. Higher variance features increase the probability that a model will capture the dataset's structure \cite{Hamel2012a}. Through this graph, an analyst can understand which features are important to the overall SOM structure, allowing them to examine predictions at a local level based on globally important features.

Figure \ref{fig: nsl-local} shows the local explanations for a prediction, where each feature on the y-axis {has a value representing distance from its respective BMU value} (See Section \ref{X-SOM}). {In this example,} we can see the features {with the largest impact on a prediction: duration, dst bytes, and src bytes.} {These features were the closest to the BMU, and they played a large role in computing the predicted value.} Seeing the specific features that influence predictions {provides insight about samples labeled as malicious or benign and can further help operators determine the reason of incorrect predictions.} These features can also be further investigated with feature value heat maps. 

\subsubsection{Unified Distance Matrix (U-Matrix)} The U-Matrix is a visualization of the distances between neighboring SOM units. With distances shown as a color gradient, units far apart will create dark boundaries while areas with similar units will be lighter. This can visually represent the natural clusters of input data. To enhance the standard U-Matrix, the starburst model uses connected component lines of nodes overlaid on the matrix to better represent clusters \cite{Hamel2012}. For a labeled data set, the user is able to visualize each BMU along with the associated label. Figure \ref{fig: nsl-U-Matrix} shows clear clusters with boundaries separating malicious (1) and benign (0) behavior. 
Using this information a security analyst can investigate more visualizations and feature importance values to gain an understanding about why certain malicious network activities are being grouped together.
    
\subsubsection{Feature Value Heat Map}
\begin{figure*}
     \begin{subfigure}[b]{0.38\textwidth}
         \centering
         \includegraphics[width=\textwidth]{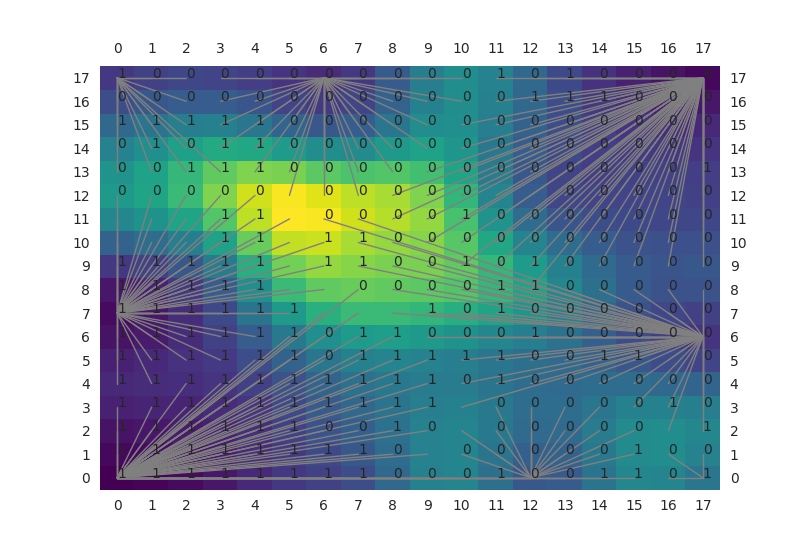}
         \caption{NSL-KDD Starburst U-Matrix}
         \label{fig: nsl-U-Matrix}
     \end{subfigure}
     \hfill
     \begin{subfigure}[b]{0.3\textwidth}
         \centering
         \includegraphics[width=\textwidth]{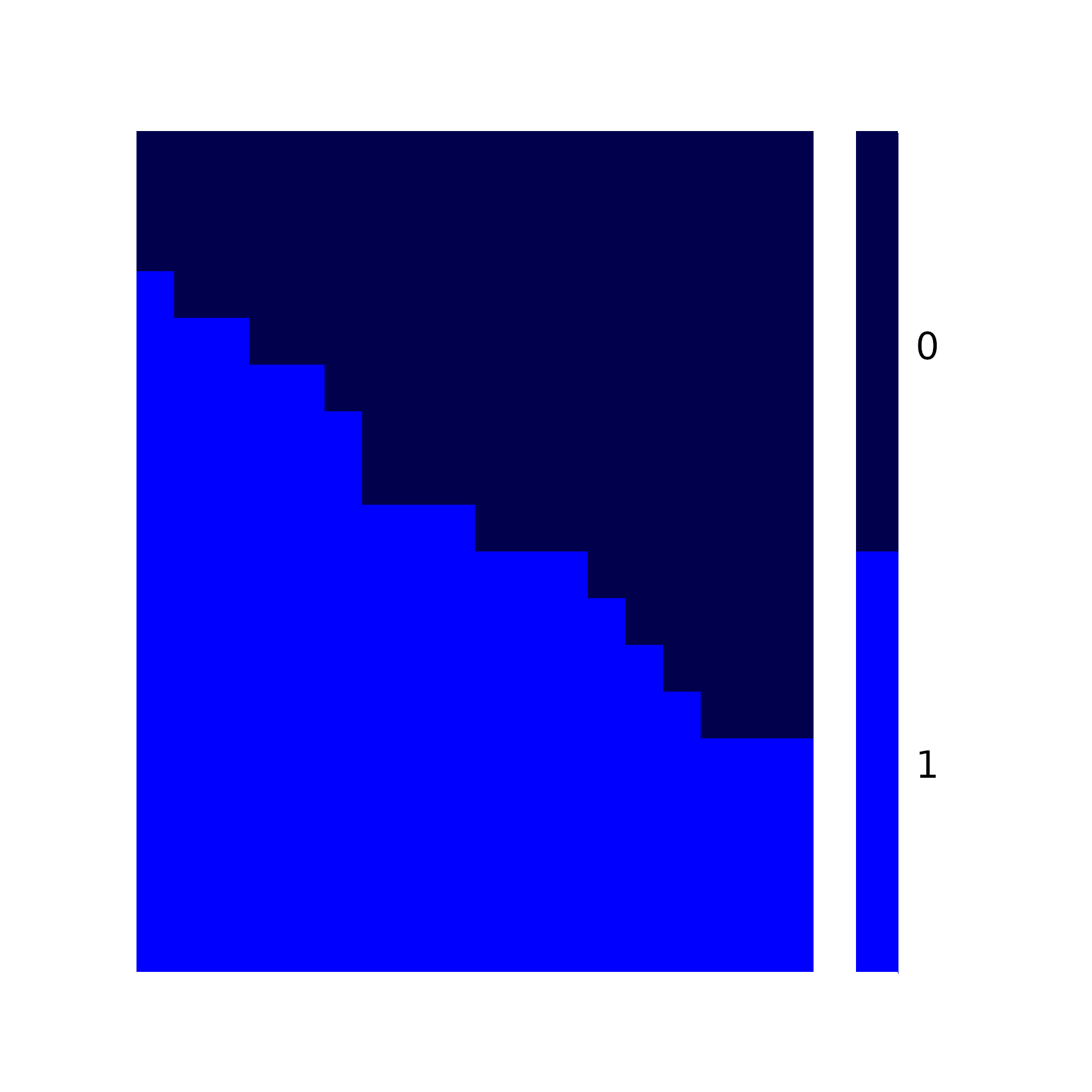}
         \caption{NSL-KDD K-means Clustering Map}
         \label{fig: nsl-K-means}
     \end{subfigure}
     \hfill
     \begin{subfigure}[b]{0.3\textwidth}
         \centering
         \includegraphics[width=\textwidth]{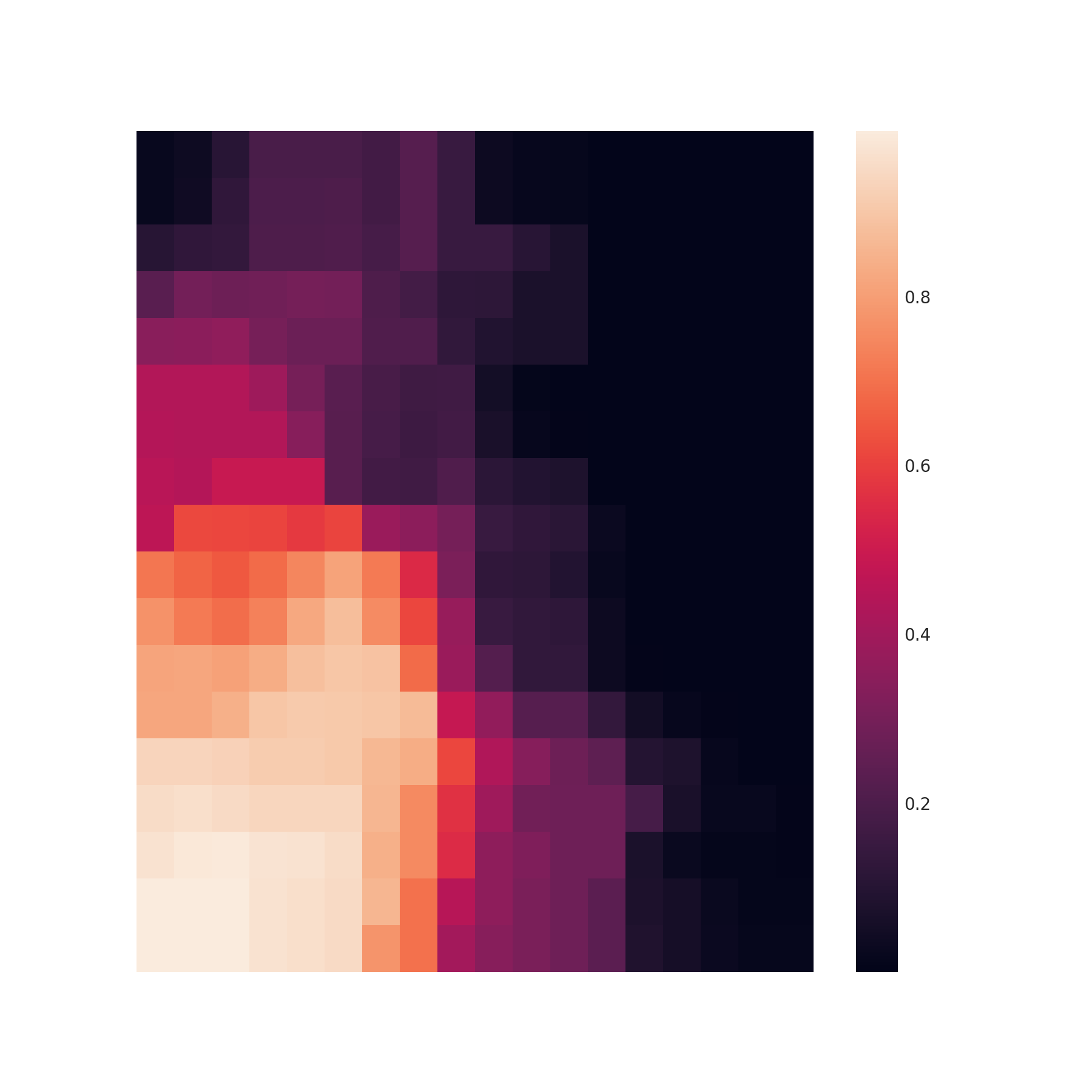}
         \caption{Dst byte Feature Map}
         \label{fig: nsl-Feature}
     \end{subfigure}
     \begin{subfigure}[b]{0.38\textwidth}
         \hspace{80pt}
         \subcaptionbox{CIC-IDS-2017 Starburst U-Matrix\label{fig: cic-U-Matrix}}{%
         \includegraphics[width=\textwidth]{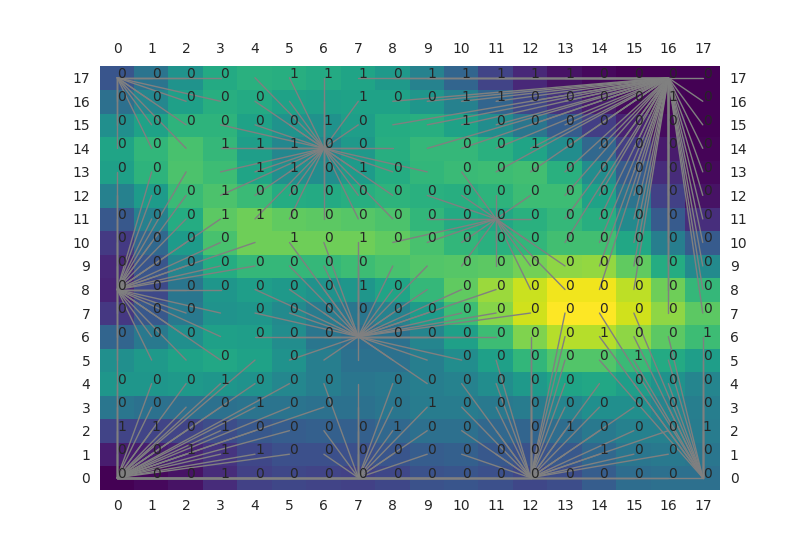}
         }
     \end{subfigure}
     \begin{subfigure}[b]{0.3\textwidth}
        \hspace{80pt}
        \subcaptionbox{Flow bytes/s Feature Map\label{fig: cic-Feature}}{%
        \includegraphics[width=\textwidth]{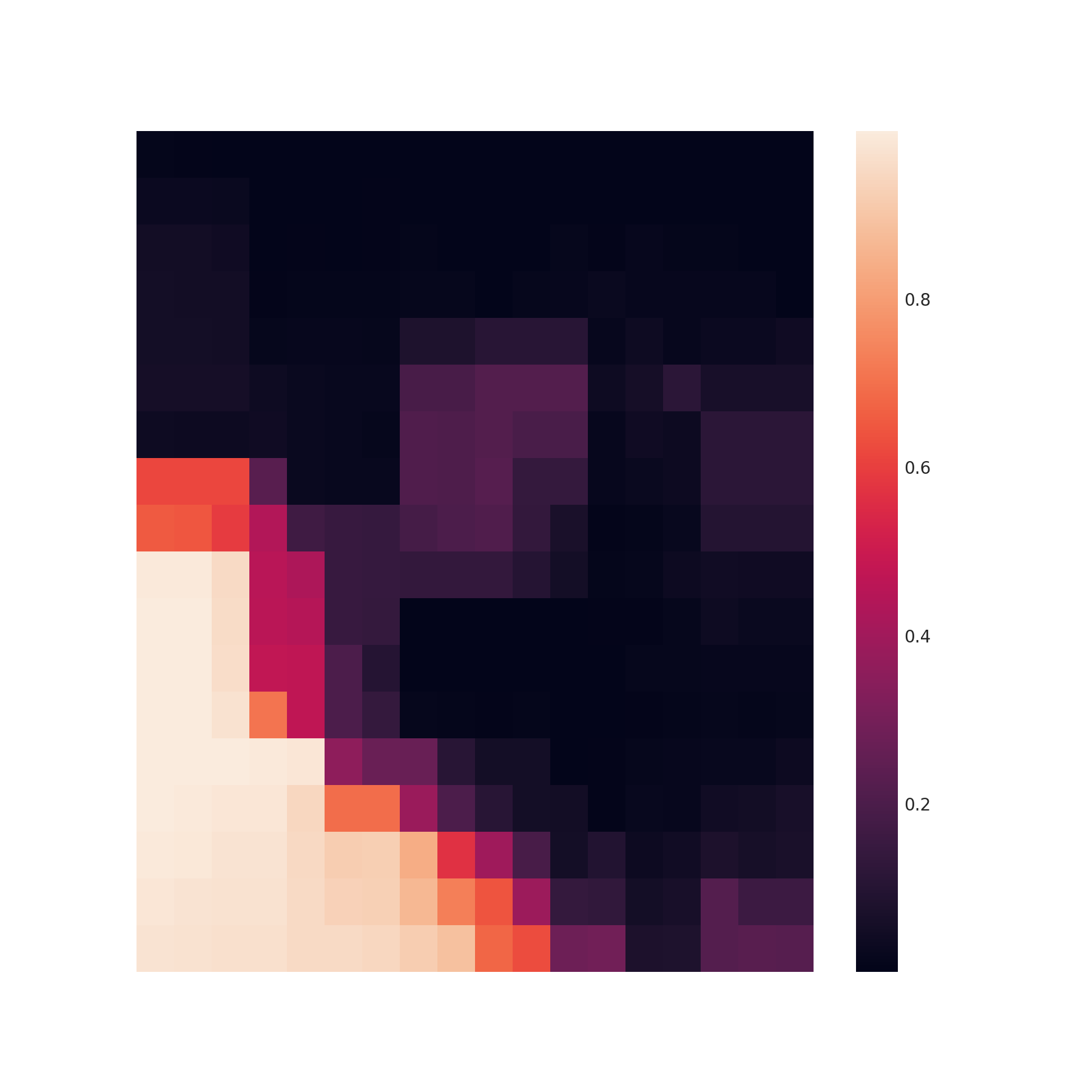}
        }
         
     \end{subfigure}
        \caption{(a)(d)The Starburst U-Matrix shows both the most common label for each node and the clusters the SOM has learned. Darker areas represent units that are close Euclidean Distance-wise. Notably, we can see a clear divide between classes on the NSL-KDD dataset as represented in the figure. (b)(e) K-means clustering can be used as a simplified view of where labels appear on the SOM. In this model's iteration, anomalous traffic is mostly grouped on the bottom of the SOM.(c) The feature value heatmap displays the value of a specific feature on each unit in the SOM. Lighter values represent units with values closer to 1, while darker values show values closer to 0. The `dst byte' example shows that the bottom `anomalous' cluster values higher values. }
        
\end{figure*}

{Heat maps show general trends that a feature has on the entire SOM model. SOM feature values are represented from 0 to 1, and the heat maps denote this with darker and lighter values, respectively. An example feature value heat map can be found in Figure \ref{fig: nsl-Feature}. In this example, the `dst bytes' features has a cluster of higher values in the bottom-left corner, while the rest of the SOM consists of lower values. Users can use this information to form conclusions about the model. Feature value maps are more powerful when multiple are viewed at a time. In addition, the U-Matrix or K-means clustering charts can then be referenced to make general decisions about the model. The heat maps work well as a fine-grained global explanation that helps users to understand the overall model.} 
    

\section{Experimental Results and Evaluation}
\label{Results}

Our SOM based X-IDS was evaluated on both \textit{explanation generation} and \textit{traditional accuracy tests}. Experiments were run using the aforementioned datasets (See Section \ref{data}), which was used to train two 18x18 SOMs. The training process was completed in 1000 iterations over the dataset. After 1000 iterations, there was no significant improvement in evaluation metrics. In fact, while training on the CIC-IDS-2017 dataset, the SOM model performance began to degrade as a result of over-fitting. 

\subsection{Model Explainability}
The results for the NSL-KDD dataset can be found in Figures \ref{fig: nsl-local} and \ref{fig: nsl-global}. The local explanation example shows that the most important features for its prediction were `Duration', `Destination (dst) bytes', and `Source (src) bytes'. The remaining features, `Service (srv) count', `Count', and `Destination (dst) host count' are considered less significant because of their distance from the BMU. Two of the important features coincide with the Global Feature Significance graph. This trend continues when testing on many different test samples. The most important global features are frequently at the forefront for local significance. Similarly to NSL-KDD, CIC-IDS-2017 follows this trend. Many of the top, globally selected features also play a more important role in the local predictions.

The next set of explainability techniques has been data-mined from the trained SOM. Figures \ref{fig: nsl-U-Matrix} and \ref{fig: cic-U-Matrix} show the generated Starburst U-Matrix for NSL-KDD and CIC-IDS-2017, respectively. The SOM algorithm was able to separate benign clusters from malicious clusters in the map created from NSL-KDD dataset. The bottom-left corner is primarily malicious samples, while the top-right corner contains mostly benign samples. Additionally, the clusters marked by the starbursts' origins mostly represent one label. On the other hand, the CIC-IDS-2017 map has not separated the labels sufficiently. Most of the labels present in the figure are benign (0) with very few malicious labels (1). CIC-IDS-2017 is an unbalanced dataset, with about 70\% of samples being benign and 30\% of samples as malicious. This class imbalance causes the SOM to be trained on more benign samples than malicious.

For a simplified label separation, users can visualize a K-means clustering interpretation in Figure \ref{fig: nsl-K-means}. This figure helps to explain which clusters the benign and malicious traffic are grouped in. The NSL-KDD K-means graph is similar to the computed U-matrix. However, the CIC-IDS-2017 K-means cluster graph was unable to form accurate clusters. As mentioned above, there were few units that were labeled malicious (1), and the K-means clustering algorithm chosen was unable to create a meaningful separation.

Lastly, the feature value heatmaps are generated for each feature of the dataset. The examples chosen were the most significant features for each dataset: `destination (dst) bytes' and `flow bytes/s'. On their own, they can be used to see general training trends for each feature. In Figures \ref{fig: nsl-Feature} and \ref{fig: cic-Feature}, we can see that each of these features have higher values in the bottom-left units and lower values elsewhere. {Users will be able to build a mental model of the SOM when visualized in conjunction with the features maps.} For example, `destination (dst) byte', `duration', and `source (src) byte' all have higher values in the malicious section of the map. One may conclude that when these values are all close to one, the sample is more than likely malicious.

\begin{table}[!htb]
    \small
    \setlength{\tabcolsep}{8pt}
\begin{tabularx}{\linewidth}{@{} c c c c c c@{}}
        \toprule
\thead{Dataset}  & \thead{F1} & \thead{Precision} & \thead{Recall} & \thead{FPR} & \thead{FNR}\\
         \midrule
        NSL-KDD & 91.0\% & 91.0\%  & 91.4\% & 9.4\% & 8.0\%\\
        CIC-IDS-2017 & 80.0\% & 77.4\% & 81.8\% & 22.5\% & 4.5\%\\

        \bottomrule
    \end{tabularx}
    \caption{Accuracy Metrics for NSL-KDD and CIC-IDS-2017}
    \label{table:accuracy}
\end{table}


\subsection{Accuracy}

When creating an IDS, accuracy is an important metric to consider. Table \ref{table:accuracy} shows the accuracy metrics obtained for both the NSL-KDD and CIC-IDS-2017 datasets. The accuracy of the NSL-KDD evaluation can be attributed to the separation of benign and malicious traffic, as mentioned above. 
The accuracy of CIC-IDS-2017, however, is much lower. The U-matrix shows that not many units are labeled as malicious. Interestingly, only 14\% of the units are labeled as malicious, which means that 77.4\% of malicious samples are similar to that small subset of units.

The results from the explainability and accuracy experiments show that it is possible to create explainable and relatively accurate SOM based X-IDS. The visualization techniques used can give users an understanding of the model and can empower security analysts to make their own predictions, similar to the model. A 91\% F1-score can be attributed to the clear separation the model makes between malicious and benign samples. We believe that this can be further improved with more complex SOM algorithms. 

\section{Conclusion and Future Work}\label{Conclusion}

In this paper, we created a proof of concept SOM based X-IDS implementation. The implementation was able to produce robust, explainable figures describing the SOM model. Explainability was demonstrated using various forms of visualization including feature significance, U-matrices, and feature heatmaps. Through these, users are able to create their own conclusions about how the model works and makes predictions. Additionally, accuracy was tested using the NSL-KDD and CIC-IDS-2017 datasets. The SOM implementation was able to achieve accuracies of 91\% and 80\%, respectively. Potential future works will include analysing the explainability and accuracy of Growing SOMs or Growing Hierarchical SOMs. These studies will aim to increase the accuracy of SOMs while simultaneously decreasing false positives and false negatives. In addition, explanations can be improved by surveying security analysts to discover the most useful visualizations and feedback. We will continue to use and improve our architecture to create the state-of-the art in X-IDS.

\section{Acknowledgment}
This work by Mississippi State University was financially supported by the U.S. Department of Defense (DoD) High Performance Computing Modernization Program, through the US Army Engineering Research and Develop Center (ERDC) (\#W912HZ-21-C0058). The views and conclusions contained herein are those of the authors and should not be interpreted as necessarily representing the official policies or endorsements, either expressed or implied, of the U.S. Army ERDC or the U.S. DoD.

\bibliographystyle{unsrt}
\bibliography{refs}

\begin{thebibliography}{10}

\bibitem{Marshan}
Alaa Marshan.
\newblock Artificial intelligence: Explainability, ethical issues and bias.
\newblock {\em Annals of Robotics and Automation}, pages 034--037, 08 2021.

\bibitem{raytheon_2017}
Raytheon.
\newblock Cyber security operations center (csoc), 2017.

\bibitem{belouch2018performance}
Mustapha Belouch, Salah El~Hadaj, and Mohamed Idhammad.
\newblock Performance evaluation of intrusion detection based on machine
  learning using apache spark.
\newblock {\em Procedia Computer Science}, 127:1--6, 2018.

\bibitem{wu2010use}
Shelly~Xiaonan Wu and Wolfgang Banzhaf.
\newblock The use of computational intelligence in intrusion detection systems:
  A review.
\newblock {\em Applied soft computing}, 10(1):1--35, 2010.

\bibitem{lipton2018mythos}
Zachary~C Lipton.
\newblock The mythos of model interpretability: In machine learning, the
  concept of interpretability is both important and slippery.
\newblock {\em Queue}, 16(3):31--57, 2018.

\bibitem{subashxids}
Subash Neupane, Jesse Ables, William Anderson, Sudip Mittal, Shahram Rahimi,
  Ioana Banicescu, and Maria Seale.
\newblock Explainable intrusion detection systems (x-ids): A survey of current
  methods, challenges, and opportunities, 2022.

\bibitem{gunning2019darpa}
David Gunning and David Aha.
\newblock Darpa’s explainable artificial intelligence (xai) program.
\newblock {\em AI Magazine}, 40(2):44--58, 2019.

\bibitem{kohonen1982self}
Teuvo Kohonen.
\newblock Self-organized formation of topologically correct feature maps.
\newblock {\em Biological cybernetics}, 43(1):59--69, 1982.

\bibitem{denning1987intrusion}
Dorothy~E Denning.
\newblock An intrusion-detection model.
\newblock {\em IEEE Transactions on software engineering}, (2):222--232, 1987.

\bibitem{bace2001intrusion}
Rebecca~Gurley Bace, Peter Mell, et~al.
\newblock Intrusion detection systems, 2001.

\bibitem{mcdole2021deep}
Andrew McDole, Maanak Gupta, Mahmoud Abdelsalam, Sudip Mittal, and Mamoun
  Alazab.
\newblock Deep learning techniques for behavioural malware analysis in cloud
  iaas.
\newblock In {\em Malware Analysis using Artificial Intelligence and Deep
  Learning}. Springer, 2021.

\bibitem{LetouHIDS}
Kopelo Letou, Dhruwajita Devi, and Yumnam Jayanta.
\newblock Host-based intrusion detection and prevention system (hidps).
\newblock {\em International Journal of Computer Applications}, 69:28--33, 05
  2013.

\bibitem{Mukherjee1994NetworkID}
Biswanath Mukherjee, Todd~L. Heberlein, and Karl~N. Levitt.
\newblock Network intrusion detection.
\newblock {\em IEEE Network}, 8:26--41, 1994.

\bibitem{sharma2014evolution}
Ashu Sharma and Sanjay~Kumar Sahay.
\newblock Evolution and detection of polymorphic and metamorphic malwares: A
  survey.
\newblock {\em arXiv preprint arXiv:1406.7061}, 2014.

\bibitem{Chandola2009AnomalyDA}
Varun Chandola, Arindam Banerjee, and Vipin Kumar.
\newblock Anomaly detection: A survey.
\newblock {\em ACM Comput. Surv.}, 41:15:1--15:58, 2009.

\bibitem{mcdole2020analyzing}
Andrew McDole, Mahmoud Abdelsalam, Maanak Gupta, and Sudip Mittal.
\newblock Analyzing cnn based behavioural malware detection techniques on cloud
  iaas.
\newblock In {\em International Conference on Cloud Computing}, pages 64--79.
  Springer, 2020.

\bibitem{szczepanski2020achieving}
Mateusz Szczepa{\'n}ski, Micha{\l} Chora{\'s}, Marek Pawlicki, and Rafa{\l}
  Kozik.
\newblock Achieving explainability of intrusion detection system by hybrid
  oracle-explainer approach.
\newblock In {\em 2020 International Joint Conference on Neural Networks
  (IJCNN)}, pages 1--8. IEEE, 2020.

\bibitem{pang2021explainable}
Guansong Pang, Choubo Ding, Chunhua Shen, and Anton van~den Hengel.
\newblock Explainable deep few-shot anomaly detection with deviation networks.
\newblock {\em arXiv preprint arXiv:2108.00462}, 2021.

\bibitem{subba2015intrusion}
Basant Subba, Santosh Biswas, and Sushanta Karmakar.
\newblock Intrusion detection systems using linear discriminant analysis and
  logistic regression.
\newblock In {\em 2015 Annual IEEE India Conference (INDICON)}, pages 1--6.
  IEEE, 2015.

\bibitem{mahbooba2021explainable}
Basim Mahbooba, Mohan Timilsina, Radhya Sahal, and Martin Serrano.
\newblock Explainable artificial intelligence (xai) to enhance trust management
  in intrusion detection systems using decision tree model.
\newblock {\em Complexity}, 2021, 2021.

\bibitem{langin2011annabell}
Chet Langin, Michael Wainer, and Shahram Rahimi.
\newblock Annabell island: a 3d color hexagonal som for visual intrusion
  detection.
\newblock {\em Internation Journal of Computer Science and Information
  Security}, 9(1):1--7, 2011.

\bibitem{Liu2008IsolationF}
Fei~Tony Liu, Kai~Ming Ting, and Zhi-Hua Zhou.
\newblock Isolation forest.
\newblock {\em 2008 Eighth IEEE International Conference on Data Mining}, pages
  413--422, 2008.

\bibitem{Schlkopf1999SupportVM}
Bernhard Sch{\"o}lkopf, Robert~C. Williamson, Alex Smola, John Shawe-Taylor,
  and John~C. Platt.
\newblock Support vector method for novelty detection.
\newblock In {\em NIPS}, 1999.

\bibitem{ZhangNN}
G.P. Zhang.
\newblock Neural networks for classification: a survey.
\newblock {\em IEEE Transactions on Systems, Man, and Cybernetics, Part C
  (Applications and Reviews)}, 30(4):451--462, 2000.

\bibitem{moore1988explanation}
Johanna~D Moore and William~R Swartout.
\newblock Explanation in expert systemss: A survey.
\newblock Technical report, University of Southern California Marina Del Rey
  Information Sciences Inst, 1988.

\bibitem{shortliffe1974mycin}
Edward~Hance Shortliffe.
\newblock Mycin: a rule-based computer program for advising physicians
  regarding antimicrobial therapy selection.
\newblock Technical report, Stanford Univ Calif Dept of Computer Science, 1974.

\bibitem{holzinger2017we}
Andreas Holzinger, Chris Biemann, Constantinos~S Pattichis, and Douglas~B Kell.
\newblock What do we need to build explainable ai systems for the medical
  domain?
\newblock {\em arXiv preprint arXiv:1712.09923}, 2017.

\bibitem{lindsay2020explainable}
Leeanne Lindsay, Sonya Coleman, Dermot Kerr, Brian Taylor, and Anne Moorhead.
\newblock Explainable artificial intelligence for falls prediction.
\newblock In {\em International Conference on Advances in Computing and Data
  Sciences}, pages 76--84. Springer, 2020.

\bibitem{chun2021study}
Ye~Eun Chun, Se~Bin Kim, Ja~Yun Lee, and Ji~Hwan Woo.
\newblock Study on credit rating model using explainable ai.
\newblock {\em The Korean Data \& Information Science Society}, 32(2):283--295,
  2021.

\bibitem{han2019joint}
Miseon Han and Jeongtae Kim.
\newblock Joint banknote recognition and counterfeit detection using
  explainable artificial intelligence.
\newblock {\em Sensors}, 19(16):3607, 2019.

\bibitem{darpa2016broad}
DARPA.
\newblock Broad agency announcement explainable artificial intelligence (xai).
\newblock {\em DARPA-BAA-16-53}, pages 7--8, 2016.

\bibitem{ribeiro2016should}
Marco~Tulio Ribeiro, Sameer Singh, and Carlos Guestrin.
\newblock "{W}hy should i trust you?" explaining the predictions of any
  classifier.
\newblock In {\em Proceedings of the 22nd ACM SIGKDD international conference
  on knowledge discovery and data mining}, pages 1135--1144, 2016.

\bibitem{lundberg2017unified}
Scott~M Lundberg and Su-In Lee.
\newblock A unified approach to interpreting model predictions.
\newblock {\em Advances in neural information processing systems}, 30, 2017.

\bibitem{binder2016layer}
Alexander Binder, Gr{\'e}goire Montavon, Sebastian Lapuschkin, Klaus-Robert
  M{\"u}ller, and Wojciech Samek.
\newblock Layer-wise relevance propagation for neural networks with local
  renormalization layers.
\newblock In {\em International Conference on Artificial Neural Networks},
  pages 63--71. Springer, 2016.

\bibitem{islam2019domain}
Sheikh~Rabiul Islam, William Eberle, Sheikh~K Ghafoor, Ambareen Siraj, and Mike
  Rogers.
\newblock Domain knowledge aided explainable artificial intelligence for
  intrusion detection and response.
\newblock {\em arXiv preprint arXiv:1911.09853}, 2019.

\bibitem{wu2020feature}
Chunyuan Wu, Aijuan Qian, Xiaoju Dong, and Yanling Zhang.
\newblock Feature-oriented design of visual analytics system for interpretable
  deep learning based intrusion detection.
\newblock In {\em 2020 International Symposium on Theoretical Aspects of
  Software Engineering (TASE)}, pages 73--80. IEEE, 2020.

\bibitem{oja1999kohonen}
Erkki Oja and Samuel Kaski.
\newblock {\em Kohonen maps}.
\newblock Elsevier, 1999.

\bibitem{guthikonda2005kohonen}
Shyam~M Guthikonda.
\newblock Kohonen self-organizing maps.
\newblock {\em Wittenberg University}, 98, 2005.

\bibitem{kohonen2007kohonen}
Teuvo Kohonen and Timo Honkela.
\newblock Kohonen network.
\newblock {\em Scholarpedia}, 2(1):1568, 2007.

\bibitem{Ong1999DataMU}
Jason Ong and Syed Muhammad~Raza Abidi.
\newblock Data mining using self-organizing kohonen maps: A technique for
  effective data clustering \& visualization.
\newblock In {\em IC-AI}, 1999.

\bibitem{pachghare2009intrusion}
VK~Pachghare, Parag Kulkarni, and Deven~M Nikam.
\newblock Intrusion detection system using self organizing maps.
\newblock In {\em 2009 International Conference on Intelligent Agent \&
  Multi-Agent Systems}, pages 1--5. IEEE, 2009.

\bibitem{Breard2017}
Gregory Breard.
\newblock Evaluating self-organizing map quality measures as convergence
  criteria.
\newblock 2017.

\bibitem{liu2018scalable}
Yao Liu, Jun Sun, Qing Yao, Su~Wang, Kai Zheng, and Yan Liu.
\newblock A scalable heterogeneous parallel som based on mpi/cuda.
\newblock In {\em Asian Conference on Machine Learning}, pages 264--279. PMLR,
  2018.

\bibitem{lichodzijewski2002host}
Peter Lichodzijewski, A~Nur Zincir-Heywood, and Malcolm~I Heywood.
\newblock Host-based intrusion detection using self-organizing maps.
\newblock In {\em Proceedings of the 2002 International Joint Conference on
  Neural Networks. IJCNN'02 (Cat. No. 02CH37290)}, volume~2, pages 1714--1719.
  IEEE, 2002.

\bibitem{de2015implementation}
Emiro De~la Hoz, Andr{\'e}s Ortiz~Garc{\'\i}a, Julio Ortega~Lopera,
  Eduardo~Miguel De~La Hoz~Correa, and Fabio~Enrique Mendoza~Palechor.
\newblock Implementation of an intrusion detection system based on self
  organizing map.
\newblock 2015.

\bibitem{rhodes2000multiple}
Brandon~Craig Rhodes, James~A Mahaffey, and James~D Cannady.
\newblock Multiple self-organizing maps for intrusion detection.
\newblock In {\em Proceedings of the 23rd national information systems security
  conference}, pages 16--19. MD Press Baltimore, 2000.

\bibitem{albayrak2005combining}
Sahin Albayrak, Christian Scheel, Dragan Milosevic, and Achim Muller.
\newblock Combining self-organizing map algorithms for robust and scalable
  intrusion detection.
\newblock In {\em International Conference on Computational Intelligence for
  Modelling, Control and Automation and International Conference on Intelligent
  Agents, Web Technologies and Internet Commerce (CIMCA-IAWTIC'06)}, volume~2,
  pages 123--130. IEEE, 2005.

\bibitem{ultsch-kohonens-data-analysis-1990}
Alfred Ultsch and H.~Peter Siemon.
\newblock Kohonen's self organizing feature maps for exploratory data analysis.
\newblock In {\em Proceedings of the International Neural Network Conference
  (INNC-90), Paris, France, July 9–13, 1990}, volume~1, pages 305--308.
  Kluwer Academic Press, 1990.

\bibitem{Wickramasinghe2021}
Chathurika~S. Wickramasinghe, Kasun Amarasinghe, Daniel~L. Marino, Craig
  Rieger, and Milos Manic.
\newblock Explainable unsupervised machine learning for cyber-physical systems.
\newblock {\em IEEE Access}, 9:131824--131843, 2021.

\bibitem{Tavallaee2009}
Mahbod Tavallaee, Ebrahim Bagheri, Wei Lu, and Ali~A Ghorbani.
\newblock A detailed analysis of the kdd cup 99 data set.
\newblock pages 1--6, 2009.

\bibitem{Panigrahi2018}
Ranjit Panigrahi and Samarjeet Borah.
\newblock A detailed analysis of cicids2017 dataset for designing intrusion
  detection systems.
\newblock {\em International Journal of Engineering \& Technology}, 7:479--482,
  3 2018.

\bibitem{Hamel2012a}
Lutz Hamel and Chris Brown.
\newblock Bayesian probability approach to feature significance for infrared
  spectra of bacteria.
\newblock {\em Applied Spectroscopy}, 66:48--59, 1 2012.

\bibitem{yuan2018implementation}
Li~Yuan.
\newblock {\em Implementation of self-organizing maps with Python}.
\newblock University of Rhode Island, 2018.

\bibitem{Kohonen1998}
Teuvo Kohonen.
\newblock The self-organizing map.
\newblock {\em Neurocomputing}, 21:1--6, 1998.

\bibitem{Lampinen1992}
Jouko Lampinen and Erkki Oja.
\newblock Clustering properties of hierarchical self-organizing maps.
\newblock {\em Journal of Mathematical Imaging and Vision}, 2:261--272, 1992.

\bibitem{Hamel2016}
Lutz Hamel.
\newblock Som quality measures: An efficient statistical approach.
\newblock volume 428, pages 49--59. Springer Verlag, 2016.

\bibitem{Tatoian2018}
Self-organizing map convergence.
\newblock {\em Int. J. Serv. Sci. Manag. Eng. Technol.}, 9:61--84, 4 2018.

\bibitem{wickramasinghe2021explainable}
Chathurika~S Wickramasinghe, Kasun Amarasinghe, Daniel~L Marino, Craig Rieger,
  and Milos Manic.
\newblock Explainable unsupervised machine learning for cyber-physical systems.
\newblock {\em IEEE Access}, 9:131824--131843, 2021.

\bibitem{Hamel2012}
Lutz Hamel and Chris Brown.
\newblock Improved interpretability of the unified distance matrix with
  connected components.
\newblock {\em 7th International Conference on Data Mining (DMIN'11)}, 4 2012.

\end{thebibliography}

\end{document}